\documentclass[a4paper]{article}
\RequirePackage[intlimits,sumlimits,namelimits,centertags]{amsmath}%
\usepackage[verbose=true,letterpaper]{geometry}
\AtBeginDocument{
	\newgeometry{
		textheight=9in,
		textwidth=6.5in,
		top=1in,
		headheight=14pt,
		headsep=25pt,
		footskip=30pt
	}
}
\usepackage{fancyhdr}
\fancyheadoffset{0pt}
\def\keywordname{{\bfseries \emph Keywords}}%
\def\keywords#1{\par\addvspace\medskipamount{\rightskip=0pt plus1cm
		\def\and{\ifhmode\unskip\nobreak\fi\ $\cdot$
		}\noindent\keywordname\enspace\ignorespaces#1\par}}

\usepackage[intlimits,sumlimits,namelimits,centertags]{amsmath}
\usepackage{natbib}
\usepackage{hyperref}
\usepackage{multirow}
\usepackage{color}
\usepackage{subfig}
\usepackage{tikz}
\usepackage{float}
\usepackage{hyperref}
\usepackage{refstyle}
\usepackage{url}
\usepackage[affil-it]{authblk}
\usepackage[onehalfspacing]{setspace}
\usepackage{multirow}
\usetikzlibrary{calc, shapes, backgrounds}
\usepackage{amssymb}
\usepackage{pdfpages}
\usepackage[T1]{fontenc}
\usepackage{booktabs}
\usepackage{lmodern}
\usepackage{color}
\begin{document}
\title{\textbf{Subgroup identification in individual patient data meta-analysis using model-based recursive partitioning}}
\author[1]{Cynthia Huber
	\thanks{Corresponding author, Email \texttt{cynthia.huber@med.uni-goettingen.de}}}
\author[2,1]{Norbert Benda}
\author[1]{Tim Friede}

\affil[1]{Department of Medical Statistics, University Medical Center G\"ottingen, G\"ottingen, Germany}
\affil[2]{Federal Institute for Drugs and Medical Devices (BfArM), Bonn, Germany}
\date{}
\maketitle

\section*{Abstract}
Model-based recursive partitioning (MOB) can be used to identify subgroups with differing treatment effects. The detection rate of treatment-by-covariate interactions and the accuracy of identified subgroups using MOB depend strongly on the sample size.
Using data from multiple randomized controlled clinical trials can overcome the problem of too small samples. However, naively pooling data from multiple trials may result in the identification of spurious subgroups as  differences in study design, subject selection and other sources of between-trial heterogeneity are ignored. In order to account for between-trial heterogeneity in individual participant data (IPD) meta-analysis random-effect models are frequently used. Commonly,  heterogeneity in the treatment effect is modelled using random effects whereas heterogeneity in the baseline risks is modelled by either fixed effects or random effects.
In this article, we propose metaMOB, a procedure using the generalized mixed-effects model tree (GLMM tree) algorithm for subgroup identification in IPD meta-analysis. Although the application of metaMOB is potentially wider, e.g. randomized experiments with participants in social sciences or preclinical experiments in life sciences, we focus on randomized controlled clinical trials.
In a simulation study, metaMOB outperformed GLMM trees assuming a random intercept only and model-based recursive partitioning (MOB), whose algorithm is the basis for GLMM trees, with respect to the false discovery rates, accuracy of identified subgroups and accuracy of estimated treatment effect. The most robust and therefore most promising method is metaMOB with fixed effects for modelling the between-trial heterogeneity in the baseline risks. \\
\keywords{model-based recursive partitioning \and subgroup identification \and  meta-analysis \and IPD
\and treatment-by subgroup interaction \and personalized medicine }
\section{Introduction}
For precision medicine it is crucial to identify subgroups which benefit differently from a specific drug or are exposed to particular harm. When subgroups are defined by single biomarkers or combinations of biomarkers, different treatment effects in subgroups imply treatment-by-biomarker interactions. We use the term biomarker here for not just genetic biomarkers, but also for other baseline patient characteristics as e.g. demographic variables \citep{biom_def}. The biomarkers involved in treatment-by-interactions are called predictive biomarkers, whereas prognostic markers predict the course of a disease. 
Several statistical approaches for identifying subgroups with differential treatment effects were proposed, e.g. review papers of \cite{tutorial} and \cite{ondra}. For subgroup identification methods, specifically tree-based methods, one of the important factors influencing the performance is sample size as has been shown in  \cite{sies}, \cite{alemayehu} and \cite{huber}. \\
Data from one randomized controlled study are therefore often not sufficient for the identification of subgroups. The sample size can be increased by pooling data from multiple studies investigating the same treatment or intervention. Although we consider randomized controlled trials here, the interest in identifying interactions of biomarkers and interventions is potentially wider, e.g. randomized experiments with participants in social sciences or preclinical experiments in life sciences.
For instance, \cite{Patel} developed a repository of individual-participant data (IPD) consisting of 19 randomized controlled trials in order to investigate whether some patient populations suffering from low back pain benefit differently from treatment. \cite{Cuijpers2007} conducted an IPD meta-analysis consisting of sixteen studies to investigate the effect of active scheduling as behavioural treatment of depression. Analysing differential effects in subgroups was also part of their research. Another example of an IPD is the International Weight Management in Pregnancy (i-WIP) database. The i-WIP Collaborative Group collected data from 36 randomized controlled trials in order to investigate the overall and differential effects of interventions based on diet and physical activity during pregnancy, primarily on gestational weight gain and maternal and offspring composite outcomes \citep{iwip}.
Differences in the study designs, study populations, quality of the studies, choice of comparator intervention or other study-specific influences can cause heterogeneity in baseline (control group) outcomes as well as in treatment effect sizes from one study to another. We discuss these two (very different) types of heterogeneity in turn.
One approach to address the heterogeneity in the baseline is to use models with stratified intercepts which estimates a separate intercept for each trial. Alternatively, the study intercepts can be considered as random assuming a suitable distribution (e.g. normal). Whereas the approach with stratified intercept does not have to make any assumptions regarding the distribution of intercepts across studies, a (parametric) random intercept approach needs fewer parameters to be estimated \citep{Legha2018}.
Incorporating between-study heterogeneity in the treatment effect into the modelling can be achieved by using random-effect models. Various types of random-effect models have been suggested for aggregated meta-analysis. \cite{Jackson2018} give an overview of random-effect models in meta-analysis for binary outcome data. 
Between-study variations of the treatment effect are addressed by assuming the treatment effects to be normally distributed in all of the models presented by \cite{Jackson2018}. Variations of the models arise due to different assumptions regarding the heterogeneity in the baseline risks or the independence of the random treatment and random baseline effects. 
In IPD meta-analysis models one-stage approaches using generalised mixed effect models are increasingly used for the analysis \citep{Simmonds2015}. One-stage models do not need to aggregate the IPD in a first step in order to analyse them with an appropriate meta-analysis models in the second step. \cite{Legha2018} focus on different approaches regarding the baseline outcomes for modelling the between-trial heterogeneity in one-stage IPD meta-analysis models. Although, it is not the main objective of Kontopantelis, one-stage models modelling the heterogeneity in the baseline differently are also compared in \cite{onevstwo}. \\
For subgroup identification methods some approaches have been suggested in order to account for trial heterogeneity. However, these approaches are not sufficiently flexible and they use rather simplistic models in order to account for between-trial heterogeneity. Subpopulation treatment effect pattern plot for meta-analysis (Meta-STEPP) \citep{metastepps} for example is a method developed for investigating treatment heterogeneity across one continuous covariate only. Meta- STEPP accounts for variations of the treatment effect across trials by two different approaches: A fixed-effect \citep{metastepps} and a random effect \citep{metastepps2} meta-analysis approach. Nevertheless the investigation of multiple covariates at a time and therefore, the detection of possible multi-way interactions with the treatment indicator are not possible in this framework. 
For tree-based subgroup identification methods some approaches to account for between-study heterogeneity are available. Accounting for between-study heterogeneity in tree-based methods is crucial as ignoring heterogeneity can lead to the identification of spurious splitting variables and therefore spurious subgroups as shown by \cite{Sela2012}.  
For SIDES \citep{sides} and model-based recursive partitioning (MOB)\citep{mobS,mobZ} extensions have been proposed, which adapt the algorithm by \cite{Sela2012}, allowing to account for heterogeneity, see \cite{Mistry2018} and \cite{Fokkema2017}, respectively. However, both investigate a simpler model assuming heterogeneity in the baseline only. Systematic combinations of tree-based subgroup identification methods and meta-analysis models which are adequately complex and flexible have not been investigated yet.\\
In this work, we introduce metaMOB which combines commonly made assumptions in meta-analysis models with generalized linear mixed model-tree (GLMM-tree) algorithm by \cite{Fokkema2017}. We investigate metaMOB's performance compared to the original MOB to the original MOB \citep{mobS,mobZ} and to the GLMM-tree investigated in \cite{Fokkema2017} in an extensive Monte-Carlo simulation study using an structured approach \citep{Benda2010}. Furthermore, we evaluate the impact of using different assumptions on the baseline effect for metaMOB. \\
The remainder of this paper is organized as follows. In Section \ref{methods} we outline the subgroup identification method MOB and its extension GLMM-trees. Moreover, we present different mixed-models appropriate for IPD meta-analysis in Section \ref{glmms}. In Subsection \ref{meta-mob} we introduce metaMOB. In the following section we compare the performance of two variations of metaMOB to MOB and the GLMM-tree investigated by \cite{Fokkema2017} in a simulation study. 
%In Section \ref{application} we illustrate the methods on a concrete data example (not included yet!). 
Finally, we discuss advantages and limitations of the investigated methods in Section \ref{discussion}.
\section{Methods}\label{methods}
%\subsection{Model}
We consider IPD data from $k=1,\ldots K$ randomized controlled trials investigating the same experimental treatment against the same control on an outcome variable $Y$. For each patient $i$ $(i= 1, \ldots, n_k)$ of a trial $k$ the outcome $y_{ik}$, $p$ baseline covariates denoted by $x_{ik1},\ldots,x_{ikp}$ and the treatment group $t_{ik}$ are observed.
The treatment variable $T$ takes the value $1$ for patients in the experimental treatment group and $0$ for patients in the control group.\\ 
\cite{mobS} use the algorithm of MOB \citep{mobZ} for exploratory subgroup identifications.
MOB assumes that in presence of subgroups there is no single global model fitting the data well. Therefore, MOB partitions the dataset with respect to some covariates in order to improve the model fit. For MOB a model for modelling the outcome has to be defined first.
Using generalized linear models (GLM), we model the expected outcome $E(y_{ik})$ of a patient given the treatment indicator $t_{ik}$ through a linear predictor and a suitable link function $g()$ assuming that all observations are drawn from the same population. 
\begin{align}
E(y_{ik}|t_{ik})= \mu_{ik}\\ \nonumber
g(\mu_{ik})= \gamma+ \theta t_{ik} \label{GLM}
\end{align}
%Equation \ref{GLM} corresponds to a single global model.\\
Partitioning the data for improving the data fit in presence of subgroups leads to separate GLMs in each obtained subgroup $j$ ($j=1,\ldots,J$):
\begin{equation}
g(\mu_{ikj})=\gamma_j+ \theta_{j}t_{ik}
\label{M0} 
\tag{$M_0$}
\end{equation}
Since we assume that the subgroup structure is not known, it is not possible to simply estimate the parameters $\gamma_j$, the intercept term of subgroup $j$, and the treatment effect $\theta_{j}$ of subgroup $j$. 
MOB grows a tree in order to find local models (see \ref{M0}) which fit the data better.\\
MOB partitions the data when the model parameters of \ref{M0} are not stable over the considered covariates $\mathbf{X}$. The instability of the model parameters are assessed using the M-fluctuation by \cite{Zeileis2007} or a permutation test when the assumption of normally distributed partial scores may not hold. More precisely the MOB tree is obtained using the following algorithm:
\begin{enumerate}
	\item Start with all observations included in the root node
	\item Fit model \ref{M0} to all observations in the given node by estimating the model parameters via minimizing the objective function $\Psi((Y,X),\gamma,\theta)$ (e.g. the negative log-likelihood)
	\item Asses parameter instabilities with respect to $\mathbf{X_1},\ldots, \mathbf{X_p}$ by using the 
	M-fluctuation test proposed by \cite{Zeileis2007} or permutation test. 
	The tested hypotheses are
	$$H_0^{\gamma,j}: \psi_\gamma((Y,\mathbf{X},T),\hat{\boldsymbol{\gamma}},\hat{\boldsymbol{\theta}})\perp X_j, \quad j=1,\ldots,p$$ 
	$$H_0^{\theta,j}: \psi_\theta((Y,\mathbf{X},T),\hat{\boldsymbol{\gamma}},\hat{\boldsymbol{\theta}})\perp X_j, \quad j=1,\ldots,p,$$
	with  $\psi_\gamma$ and $\psi_\theta$ denoting the partial score functions of $\boldsymbol{\gamma}$ and $\boldsymbol{\theta}$, respectively.
	\item  If at least one of the $2\times p$ null hypotheses can be rejected at a pre-specified nominal level (using Bonferroni multiplicity adjustment), select covariate $X_{j*}$ with the lowest p-value as splitting variable.
	\item Compute the split point for the chosen variable $X_{j*}$ by optimizing the sum the objective functions of the conceivable subsets.
	\item Repeat Steps  2 to 5 until none of the hypotheses in Step 3 can be rejected or some other stopping criteria (e.g. minimum number of observations in a node) is met.
\end{enumerate}

%In the following we present different variations of MOB for identifying subgroups with differential treatment effects. The methods differ in their assumptions regarding the between-study heterogeneity. Section \ref{MOB} provides a description of MOB for subgroup identification as proposed by Seibold et al. \cite{mobS}. 
Since MOB was developed to identify subgroups based on data from one single study, MOB ignores the clustered structure of the data. \cite{Fokkema2017} proposed an algorithm extending MOB in order to allow the assumption of heterogeneity between clusters.
The extended algorithm called GLMM-trees uses mixed models in order to model data from $K$ studies. In contrast to GLMs (as used in MOB) assuming independent and identically distributed observations, mixed models assume that observations between studies are independent but observations within each study are correlated. 
The algorithm of GLMM-trees allows to detect treatment-by-subgroup interactions and non-linearities in generalized linear mixed-effect models (GLMM). \cite{Fokkema2017} restricted their analyses to GLMMs with cluster-specific random intercepts fitting the following model in each partition $j$:
\begin{align}
g(\mu_{ijk})=\gamma_j+ \theta_{j}t_{ijk}+b_{0k}.
\label{M1}
\tag{$M_1$}
\end{align}
The centred random intercepts $b_k$ are normally distributed $b_k$ with expectation 0 and variance $\tau_0^2$. GLMM-trees using model \ref{M1} are referred to as MOB-RI in the following. \\
A random-intercept model allows only the baseline outcome to vary from study to study whereas the treatment effects are assumed to be the same across studies. This assumption corresponds to a common-effect meta-analysis model which is also known under the term fixed-effect meta-analysis model. The assumption of estimating the same true treatment effect in every trial is a strong assumption \citep{Jackson2018}.  Therefore, the random-effects model is more plausible. Random-effects models for meta-analyses are outlined in Subsection \ref{glmms}.  

\subsection{GLMMs for meta-analysis}\label{glmms} %Commonly used models
Heterogeneous treatment effects are usually expected in meta-analysis. However, this assumption has not been evaluated in the context of GLMM-trees so far. 
Several models for random-effect meta-analyses have been suggested. \cite{Jackson2018} examined seven models for random-effect meta-analyses with binary outcome using aggregated data. \cite{Legha2018} compare two approaches for one-stage IPD meta-analysis: random-effect models with either random or stratified intercepts. These two approaches for modelling the baseline are also considered in \cite{Jackson2018} and \cite{onevstwo}.
The random-effects model with stratified intercepts is defined as follows:

\begin{align}
&g(\mu_{ik})=\gamma_{k}+ \theta_{k}t_{ik}+\epsilon_{ik}\label{remodel_m3}\\
&\text{with }\theta_{k}=\theta+b_{1k}\nonumber\\
%&\text{with } \epsilon_{ik}\sim \mathcal{N}(0,\sigma^2_k), \quad 
&\text{and } b_{1k}\sim \mathcal{N}(0,\tau_1^2)\nonumber
%\label{M2_overall}
\end{align}
The baseline for trial $k$ is denoted by $\gamma_k$ and is assumed to fixed. The model parameter $\theta_k$ describes the treatment effect in trial $k$ which is assumed to be normally distributed with mean $\theta$ and a between-trial variance $\tau_1^2$.\\
The model for meta-analysis accounting for both heterogeneity in the treatment effect and in the baseline by using random effects is defined in Equation \ref{remodel_m2}.
\begin{align}%\label{M3_overall}
&g(\mu_{ik})=\gamma_{k}+ \theta_{k}t_{ik}+\epsilon_{ik}\label{remodel_m2}\\
&\text{with }\gamma_{k}=\gamma+b_{0k},\quad \theta_{k}=\theta+b_{1k} \nonumber\\
%&\text{with } \epsilon_{ik}\sim \mathcal{N}(0,\sigma^2_k), \quad 
&\text{and }  b_{0k}\sim \mathcal{N}(0,\tau_0^2) \quad \text{and } b_{1k}\sim \mathcal{N}(0,\tau_1^2)\nonumber
%\label{randomeffectmodel_1}
\end{align}
The centered random effects $ b_{0k}$ and $ b_{1k}$ are assumed to be normally distributed. As in most models considered in \cite{Jackson2018} we assume that the random effects $b_0$ and $b_1$ are independent.  \\

\subsection{metaMOB}\label{meta-mob}
As for MOB, metaMOB assumes that in the presence of subgroups the parameters $\theta$ and $\gamma$ of Equation \ref{remodel_m2} and \ref{remodel_m3}
may not describe the data well. Therefore, the data is partitioned in $J$ subgroups by using a tree model. 
Using the model described in Equation \ref{remodel_m2} which is analogous to the Simmonds and Higgins' model with random study-specific effects \citep{Jackson2018}, we fit the following model to each subgroup $j$ identified by the algorithms described in Section \ref{glmmtree_algo}:
\begin{align}
g(\mu_{ijk})=\gamma_j+b_{0k}+\theta_{j}t_i + b_{1k}t_i,
\label{M2}
\tag{$M_2$}
\end{align}
where $ b_{1k}\sim\mathcal{N}(0,\tau_1^2)$ and all  $ b_{1k}$ are independent. 
As for \ref{M1} we assume that the fixed effects $\gamma$ and $\theta$ are subgroup specific, also referred to as local parameters, whereas the random effect part is assumed to be global, meaning that the random effects are the same across the identified subgroups.\\
The model fitted in each partition $j$ assuming fixed baseline effects in each study is defined as:
\begin{align}
g(\mu_{ijk})=\gamma_{jk}+\theta_{j}t_i + b_{1k}t_{ijk},
\label{M3}
\tag{$M_3$}
\end{align}
where the fixed-effects $\gamma_{jk}$ and the mean treatment effect $\theta_{j}$  are assumed to be subgroup specific. For model \ref{M3} the number of parameters to be estimated increases with an increasing number of identified subgroups and increasing number of trials.\\
The method assuming \ref{M2} to be the underlying model for the tree growing procedure is called metaMOB-RI, as a random intercept is assumed for the baseline effects. The tree growing procedure using the model with a stratified intercept, \ref{M3}, is referred to as metaMOB-SI in the following.

\subsection{Algorithm}\label{glmmtree_algo}

Models $M_0$ to $M_3$ can be represented by the model equation
\begin{align}
g(\mu_{ij})= \mathbf{x*_{i}}^T
\begin{pmatrix}
\boldsymbol{\gamma_j}\\\boldsymbol{\theta_j}
\end{pmatrix}
+\mathbf{z}_i^T\boldsymbol{b},\label{general_model}
%\tag{GLMMtree}
\end{align}
where $\mathbf{b}$ is a vector of random effects, vector $\mathbf{z}_i^T$ is the $i$-th row of the design matrix $\mathbf{Z}$ for the random effects, vector $\mathbf{x*_{i}}^T$ is the $i$-th row of the design matrix $\mathbf{X}$ for the fixed effects. The coefficient vector of the fixed effects is denoted by$(
\boldsymbol{\gamma_j},\boldsymbol{\theta_j})^T$.
For model \ref{M2} e.g.  vector $\mathbf{b}$ of Equation \ref{general_model} is $\mathbf{b}=(b_{01},\ldots,b_{0K},b_{11},\ldots,b_{1K})^T$. Vector $\mathbf{z}_i$ is a vector of length $2K$, indicating to which study patient $i$ belongs. Subject $i$ enrolled in the active treatment arm of study $k$ has the value $1$ at position $k$ and $K+k$. A subject $i$ enrolled in the placebo arm of study $k$ is a unit vector with value one at position $k$. And vector $\mathbf{x*_{i}}$ of Equation \ref{general_model} is of length $2K$ with value 1 at position $k$ and the value indicating the treatment patient $i$ was assigned to at position $K+k$.\\
The vectors $\mathbf{b}$, $\mathbf{x*}_i$ and $\mathbf{z}_i$ depend on the chosen model (\ref{M0}-\ref{M3}) . The algorithm for MOB-RI, metaMOB-RI and metaMOB-SI is outlined in the following.
\begin{enumerate}
	\item Set $r=0$ and all values $\hat{\mathbf{b}}^{(r)}$ to $0$.
	\item Set $r=r+1$.  Estimate a GLM tree using $z_i^T\hat{b}^{(r-1)}$ as an offset. The random effects part is treated as offset because it is assumed to be equal across all subgroups and it is treated as known.
	\item Fit the chosen linear mixed effect models (model \ref{M1},\ref{M2} and \ref{M3}) with terminal node $j(r)$ from the GLM tree estimated in the previous step.\\ 
	Extract posterior predictions $\hat{b}_{(r)}$ from the estimated model.
	\item Repeat (2) and (3) until the log-likelihood of the model in the previous steps converges. This is usually the case, when the estimated tree does not change from previous iterations.
\end{enumerate}

\section{Simulation Study}\label{simulation}
In Section \ref{methods} we outlined four models used for the subgroup identifications procedure MOB, GLMM-tree and metaMOB. Model \ref{M0} corresponds to a simple GLM ignoring the clustered structure of the data. MOB as proposed by \cite{mobS} aims at achieving a better model fit by fitting  Model \ref{M0} to each resulting partition.
The tree-methods accounting for clustered data by assuming \ref{M1},\ref{M2} and \ref{M3} to be the underlying models for the algorithm in Section \ref{glmmtree_algo} are referred to as MOB-RI, metaMOB-RI and metaMOB-SI, respectively.
In our simulation study using the Clinical Scenario Evaluation (CSE) framework \citep{Benda2010} we compared the performance of MOB, MOB-RI, metaMOB-RI and metaMOB-SI in different IPD settings described in Section \ref{simsetting}. CSE structures the simulation study into three parts, namely assumptions, options and metrics. \cite{cse} use a different terminology for the three components. The three components are named data models, analysis models and evaluation models. Assumptions define the data generation process, options describe the methods applied to the data and metrics specify the criteria for evaluating the methods applied to the data.
The performance is assessed by different measures presented in Section \ref{opcharacteristics}. The computational details and the chosen tuning parameters for the algorithms are given in Section \ref{compdetail}.

\subsection{Assumptions: Simulation settings} \label{simsetting}
Each dataset generated for our simulation study consists of a continuous response $Y$, a binary treatment variable $T$ and 15 covariates for each subject $i = 1, \ldots,n_k$ of $k=1,\ldots,K$ trials. For simplicity, we assume that each trial has the same number of observations $n_1=\ldots =n_K$. The total number of observations is denoted by $N$. The treatment indicator $T$ is either 0 or 1, each with a probability of 0.5. Adapted  from \cite{stima} and \cite{Fokkema2017} the 15 covariates $X_1$,\ldots,$X_{15}$ are drawn from a multivariate normal distribution with $\mu_{X_1}=10, \mu_{X_2}=30,\mu_{X_4}=-40$ and $\mu_{X_5}=70$. The other means are drawn from a discrete uniform distribution on the interval $[-70,70]$. The variance of all $X_p$ is set to $\sigma_{X_p}^2=100$. Furthermore, all 15 covariates are correlated with $\rho=0.3$. 
The outcome $Y$ is generated by $y_{ik}=f(x_{ik},t_{ik})+b_{0k}+b_{1k}*t_{ik}+\epsilon_{ik}$, with $\epsilon_{ik}\sim N(0,5^2)$, $b_{0k}\sim N(0,\tau_0^2)$ and $b_{1k}\sim N(0,\tau_1^2)$. 
The values for the varied parameters are presented in Table \ref{tab:parameterssimulation} . For each setting resulting from combining these parameters 2000 datasets are generated.

\begin{table}[h]
	\begin{tabular}{p{5cm}|p{8cm}}
		Parameter& Values\\
		\hline
		Number of trials $K$ & 5, 10 \\
		
		Sample size $N$ & 200, 500, 1000  \\
		
		Heterogeneity in baseline $\tau_0$ & 0, 5, 10  \\
		
		Heterogeneity in treatment $\tau_1$ & 0, 2.5, 5, 10 \\
		
		Correlation between $b_0$ or $b_1$ and one of the $X_p$ variables&$b_0$ (or $b_1$) and all $X_p$ uncorrelated, $b_0$ (or $b_1$) correlated with one of the splitting variables (correlation r $\approx$ 0.42), $b_0$ (or $b_1$) correlated with one of the non-splitting variables (correlation r $\approx$ 0.42)  \\
		
	\end{tabular}
	\caption{Scenarios considered in the simulation study}
	\label{tab:parameterssimulation}
\end{table}

\begin{figure}
	\begin{tikzpicture}[
	scale = 1.2, transform shape, thick,
	every node/.style = {
		shape=ellipse,
		draw,
		text width=1.5cm,
		text centered,
		minimum size=10mm
	},
	grow = down,  % alignment of characters
	level 1/.style = {sibling distance=7cm},
	level 2/.style = {sibling distance=3cm}, 
	level 3/.style = {sibling distance=3.5cm}, 
	level 4/.style = {sibling distance=1cm}, 
	level distance = 1.95cm,
	level 2/.append style={level distance=3cm}
	%level1/.style ={level distance=1.95cm},
	%level2/.style ={level distance=3cm},
	]
	\
	\node[](Start) {$X_2$} 
	%child { 
	%node[] 
	%     {\Large  $X_1$}
	child {   node [label=center:\textsf{$X_1$}] (A) {}
		child { node [shape=rectangle,minimum size=23mm,label = center:\textsf{
				$\gamma_{2k}+\theta_2T$\\with $\mu_{\gamma_1}$=17.5\\and $\theta_1$=-5}] (B) {}}
		child { node [shape=rectangle,minimum size=23mm,label = center:\textsf{
				$\gamma_{2k}+\theta_2T$\\
				with $\mu_{\gamma_2}$=30\\ and $\theta_2$=0}] (C) {}}
	}
	child {   node [text = black,label = center:\textsf{$X_5$}] (D) {}
		child {node [shape=rectangle,minimum size=23mm,label = center:\textsf{$\gamma_{3k}+\theta_3T$\\
				with
				$\mu_{\gamma_3}$=30\\ and $\theta_3$=0}] (E) {}}
		child { node [shape=rectangle,,minimum size=23mm,label = center:\textsf{
				$\gamma_{4k}+\theta_4T$\\
				with
				$\mu_{\gamma_4}$=42.5\\and $\theta_4$=5}] (F) {}}	
	}
	% }
	;
	
	% Labels
	\begin{scope}[nodes = {draw = none}]
	\path (Start) -- (A) node [near start, left]  {$\leq 30$};
	\path (A)     -- (B) node [near start, left]  {$\leq 17$};
	\path (A)     -- (C) node [near start, right] {$> 17$};
	\path (Start) -- (D) node [near start, right] {$> 30$};
	\path (D)     -- (E) node [near start, left]  {$\leq63$};
	\path (D)     -- (F) node [near start, right] {$>63$};
	
	\end{scope}
	\end{tikzpicture}
	
	\caption{Data generating model adapted from \cite{stima} and \cite{Fokkema2017} for settings Sim A and Sim B. }
	\label{fig.sim1}
\end{figure}
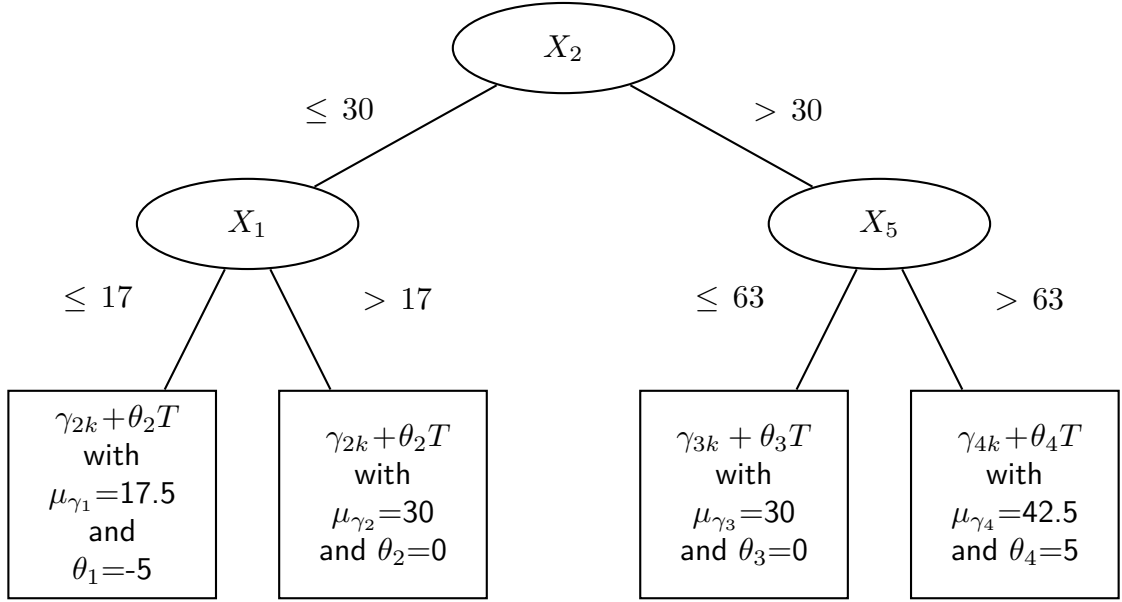

We investigated different data-generating models by varying the functional relationship of the outcome and the covariates $f(\cdot)$ as defined in Table \ref{tab:namesetting}. The Null model is used for investigating the false discovery rate. Sim A or Sim B are based on the tree structure in Figure \ref{fig.sim1} adapted from \cite{stima} and \cite{Fokkema2017}.
\begin{table}[h]
	\begin{tabular}{l|p{12cm}}
		
		Setting & Functional relationship\\
		\hline
		Null model & $f(\cdot)=0$ \\
		%\hline
		Sim A & $f(\cdot)$ is given by the tree structure presented in Figure \ref{fig.sim1} with $\gamma_{j1}=\ldots= \gamma_{jK}=\mu_{\gamma_j}$, for $j=1,\ldots,4$  \\
		%\hline
		Sim B & $f(\cdot)$ is given by the tree structure presented in Figure \ref{fig.sim1} with intercepts drawn from normal distributions: $\gamma_{1k} \sim N(17.5,\tau_{\gamma})$, $\gamma_{2k} \sim N(30,\tau_{\gamma})$,$\gamma_{3k} \sim N(17.5,\tau_{\gamma})$, $\gamma_{4k} \sim N(42.5,\tau_{\gamma})$ for $k=1,\ldots,K$ \\
		%	\hline
	\end{tabular}
	\caption{ Functional relationship of the outcome and the covariates for the simulation study}
	\label{tab:namesetting}
\end{table}

The true data-generating model for settings with $\gamma_{j1}=\ldots= \gamma_{jK}=\mu_{\gamma_j}$ is either \ref{M0}, \ref{M1} or \ref{M2} depending on the values of $\tau_0^2$ and $\tau_1^2$. Drawing the overall mean for each study within a subgroup from a normal distribution corresponds to \ref{M3} being the data generating model. Settings using Figure \ref{fig.sim1} with equal intercept across studies within one subgroup are going to be referred to as Sim A and settings with study-specific intercepts will be referred to as Sim B, respectively.   \\
Furthermore, it has to be noted that for evaluating the false discovery rate using the Null model, all $X_p$ are non-splitting variables. Therefore, $b_0$ or $b_1$ are either uncorrelated with all of covariates, or correlated with a non-splitting covariate.

\subsection{Options: Methods and their tuning parameters} \label{compdetail}
For the computation we used the software environment R \citep{Rcore}. Version 1.2-3 of the R package \texttt{partykit}\citep{partykit} was used for growing MOB described in Section \ref{methods}. For the MOB-RI and the two variations of metaMOB we used Version 0.1-2 of the \texttt{R} package \texttt{glmertree}. Since our simulations study focuses on a continuous outcome the functions used for the tree algorithms are called \texttt{lmtree} for MOB and \texttt{lmertree} for MOB-RI and metaMOB.
The stopping criteria were set to 5\% level of significance for the test in the splitting criterion and a minimum number of 20 observations in the terminal nodes. Furthermore, we set the tolerance in the argument \textit{check.conv.grad} of \texttt{lme4} (Version 1.1-21) to a five times higher value than the  default. The \texttt{glmertree} package uses \texttt{lme4} for step 3 of its algorithm (see Section \ref{glmmtree_algo}).
The mixed models fitted for MOB-RI and metaMOB using \texttt{lme4}'s default value have a poor convergence. \cite{lmeconvergence} argue that convergence issues with the max|grad| tolerance arising from the preset tolerance in the \textit{check.conv.grad} argument is the least problematic of convergence errors. Moreover, they state that in practice many researcher tend to ignore these convergence errors if no other convergence issues are present. For metaMOB and MOB-RI we use REML estimation for fitting the linear mixed effect models (Step 3 of the algorithm, see Section \ref{glmmtree_algo}).
The formulas used in the functions \texttt{lmtree} or \texttt{lmertree} for the four algorithms are given below:
\begin{itemize}
	\item MOB: \texttt{ y $\sim$ factor(trt) | x1+ ... +xp}
	\item MOB-RI: \texttt{ y $\sim$ factor(trt) | trial| x1+ ... +xp}
	\item metaMOB-RI: \texttt{ y $\sim$ factor(trt) | (1|trial)+(trt-1|trial)| x1+ ... +xp}
	\item metaMOB-SI: \texttt{ y $\sim$ factor(trial)+factor(trt) | (trt-1|trial)|| x1+ ... +xp}
\end{itemize}
\texttt{trt} is the vector including the treatment indicator, \texttt{trial} is a vector indicating to which trial an observation belongs and \texttt{x1, ...+xp} are the vectors of the $p$ covariates considered as potential predictive markers.

\subsection{Metrics: Performance criteria}\label{opcharacteristics}

The performance of the methods was assessed by different measures as false discovery rate, number of identified subgroups, tree accuracy and the correlation %of the estimated and true individual treatment effect. 
of the true and estimated treatment effects in the identified subgroups. Furthermore, we evaluate the computation time of the four algorithms and convergence problems. These performance criteria are explained in more detail below. \\
The frequency of times the algorithm identified more than one final subgroup although none was present is referred to as false discovery rate. The false discovery rate is also often referred to as Type I error rate, although strictly speaking no test is performed. The null hypothesis, that no treatment-covariate interaction is present, is rejected when a tree with at least one split is identified. The number of identified subgroups corresponds to the number of terminal nodes. The tree accuracy takes the number of identified subgroups into account and additionally evaluates the selection of the splitting covariate and the selected cut-off value. A tree is considered to be accurate if the identified tree has the correct number of terminal nodes, all splitting variables are selected correctly and when the selected cut-off values for the split denoted by $c$ are in the interval $c\pm 5$ with $5$ corresponding to the population standard deviation.

\subsection{Results}

\subsubsection{Convergence and computation time}
The GLMM-algorithm described in Section \ref{glmmtree_algo} is an iterative algorithms which stops when the difference of the log-likelihood criterion of the corresponding mixed-effect model \ref{M1}, \ref{M2} or \ref{M3} from two consecutive iterations is below a threshold. This is the case when trees of subsequent iterations do not change. MOB-RI, metaMOB-RI and metaMOB-SI always converged within 2 to 3 iterations steps within our simulation study. As threshold for the convergence criterion we used the default value of the \verb"glmertree" \texttt{R}-package, namely \verb"abstol = 0.001".\\
Convergence problems of MOB-RI and metaMOB are due to fitting linear mixed-effects regression models (\ref{M1},\ref{M2} or \ref{M3}) in each iteration step . 
The \verb"lme4" \texttt{R}-package reports two different convergence warnings for MOB-RI, metaMOB-RI and metaMOB-SI. The most frequent of the convergence warnings, more than 80\% of all the warnings in our simulations, is the warning Eager and Roy assume to be the least problematic. Another problem for fitting the models \ref{M1}, \ref{M2} and \ref{M3} in the simulation study arises due to the evaluation of the scaled gradient. However, convergence problems were rare in the simulation study, see Table \ref{tab:frq_conv}. The largest number of obtained convergence warnings across the three MOB algorithms accounting for clustered data was obtained by metaMOB-RI which uses model \ref{M2} and therefore estimates the variance component of two random effects. In practise, researches often simplify their random effect structure in the presence of convergence problems. If metaMOB-RI's underlying mixed-model is the model of choice and researches are faced with convergence problems, metaMOB-SI should be used instead. This is also recommended by \cite{onevstwo} for IPD meta-analysis one-stage models.
Further operation characteristics as for example the tree accuracy are based on cases without convergence warnings.

\begin{table}[h]	
	\centering
	\begin{tabular}{c|ccc}
		&MOB-RI  &metaMOB-RI  &metaMOB-SI  \\ 
		\hline 
		Null Model (162 Settings)& 0.02\% &1.59\%  & 0.07\%  \\ 
		%	\hline 
		Sim1 (324 Setting)& 0.05\% &1.61\%  & 0.07\% \\ 
		%	$\text{cor}(b_0,\mathbf{X})=0$ (162 Setting) & 0.05\% &1.60\%  & 0.06\% \\ 
		%	$\text{cor}(b_1,\mathbf{X})=0$ (162 Setting) & 0.05\% &1.61\%  & 0.07\% \\ 
		%	\hline 
		Sim2 (162 Settings)&  0.05\%& 1.39\% &0.06\%  \\ 
		%	\hline 
	\end{tabular} 
	\caption{Frequency of convergence problems using MOB-RI, metaMOB-RI or metaMOB-SI. For each setting 2000 simulation runs were performed.}
	\label{tab:frq_conv}
\end{table}

The computation time of MOB is the smallest, since the data partition is only performed once. The mean computation time for MOB is $\sim2$s. MOB-RI and metaMOB-RI need $\sim5$s on average for growing a tree adjusting for between-study heterogeneity and metaMOB-SI needs $\sim7$s for growing a tree. The computation times for all methods seem reasonable. 
The maximum calculation time was 215 seconds using metaMOB-RI.%for Sim 1 with b1 and X uncorrelated

\subsubsection{False discovery rate}\label{typeIerror}
\begin{figure}
	\centering
	\includegraphics[width=1\linewidth]{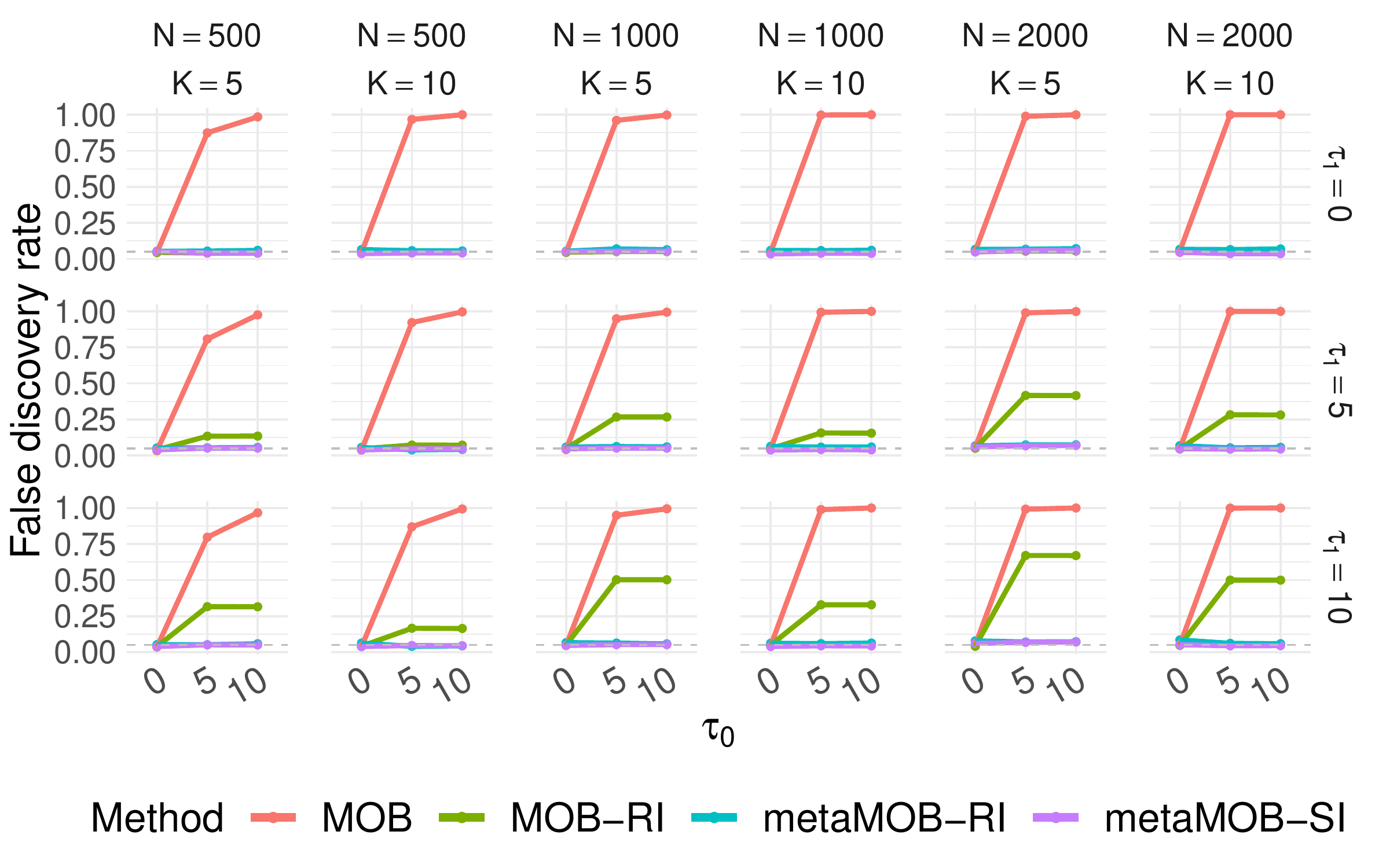}
	\caption{False discovery rate in setting with $b_0$ correlated with one of the covariates but $b_1$  uncorrelated. Dotted line at value 0.05 indicates the pre-specified level of significance used as stopping criteria in the algorithm. Rows represents the variance of the random treatment effect and columns represent sample sizes and number of trials. }
	\label{fig:type1error_b0cor_agg}
\end{figure}

\begin{figure}
	\centering
	\includegraphics[width=1\linewidth]{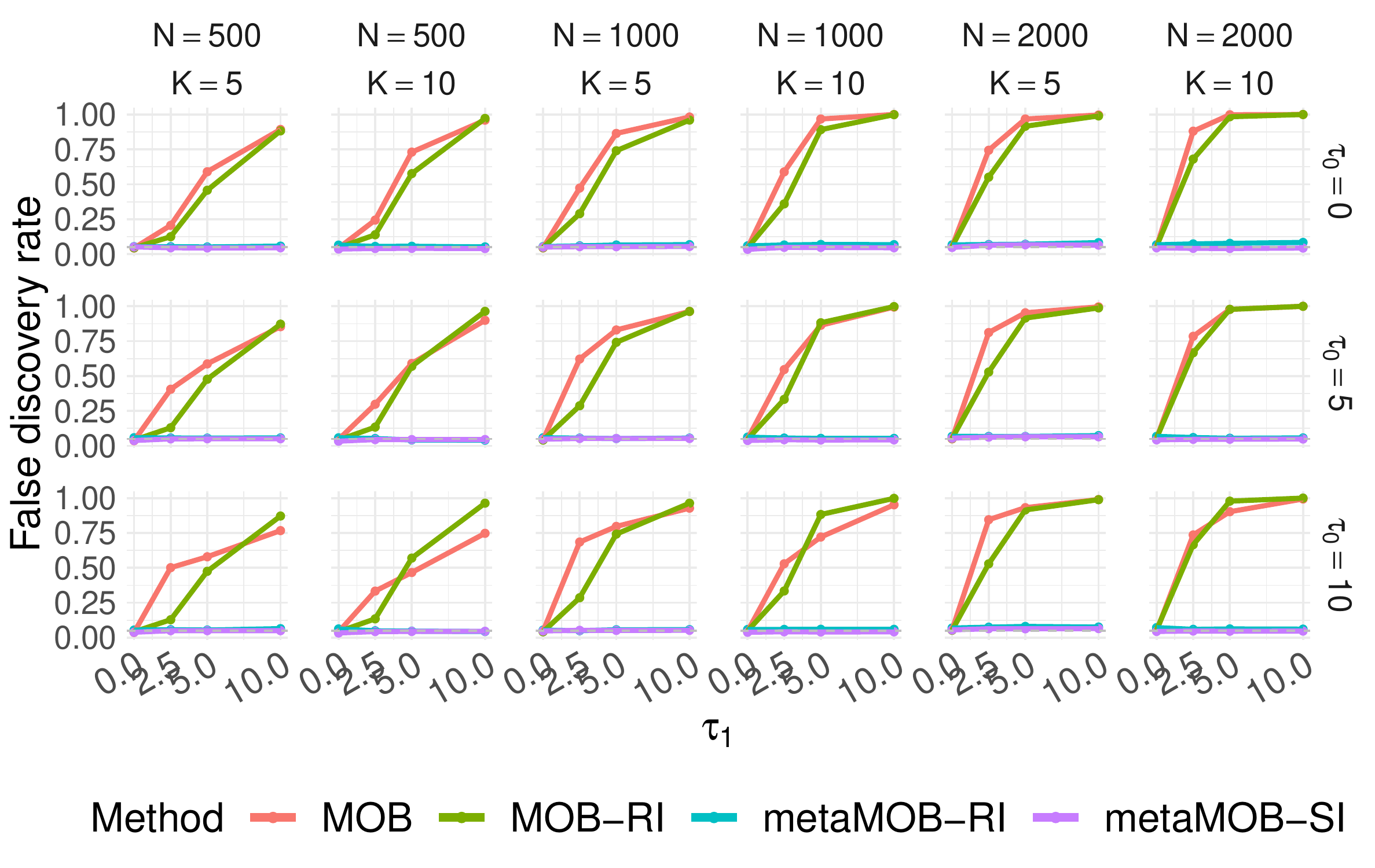}
	\caption{False discovery rate for settings with $b_1$ correlated with one of the covariates but $b_0$  uncorrelated.  Dotted line at value 0.05 indicates the pre-specified level of significance used as stopping criteria in the algorithm.}
	\label{fig:type1error_b1cor_agg_axis2}
\end{figure}

The false recovery rate is assessed by using the Null model for data generation. 
When the random  effects $b_0$ or $b_1$ are not correlated with one of the covariates the false discovery rates lie below 0.055 for all the considered settings and methods. 
False discovery rates slightly larger than 0.05 are mainly observable for settings including 2000 observations in total.
Benefits of using methods accounting for the between trial heterogeneity as MOB-RI, metaMOB-RI and metaMOB-SI are observable in the false discovery rates for settings in which the random effects $b_0$ or $b_1$ are correlated with one of the splitting candidates. 
Figure \ref{fig:type1error_b0cor_agg} shows, that  MOB's false discovery rate increases with increasing variance of the random intercept $\tau_0^2$, whereas MOB-RI, metaMOB-RI and metaMOB-SI's false discovery rates are smaller %close to 0.08
when no treatment heterogeneity between studies is present. When the variance of both random intercept and random slope are unequal to zero resulting in \ref{M2} being the true underlying model, the use of MOB-RI results in higher false discovery rates (see Figure \ref{fig:type1error_b0cor_agg}) since it does not adjust for heterogeneity in the treatment effect. A correlation of the random slope $b_1$ with one of the covariates additionally adds to spurious detections of subgroups using MOB-RI. This is depicted in Figure \ref{fig:type1error_b1cor_agg_axis2}. metaMOB-RI's and metaMOB-SI's false discovery rates are not strongly influenced by the presence of random effects or a correlation of  random effects with a splitting candidate. For the two variations of metaMOB we mainly observe false discovery rates close to 0.05.


\subsubsection{Tree accuracy}

\begin{figure}
	\centering
	\includegraphics[width=1\linewidth]{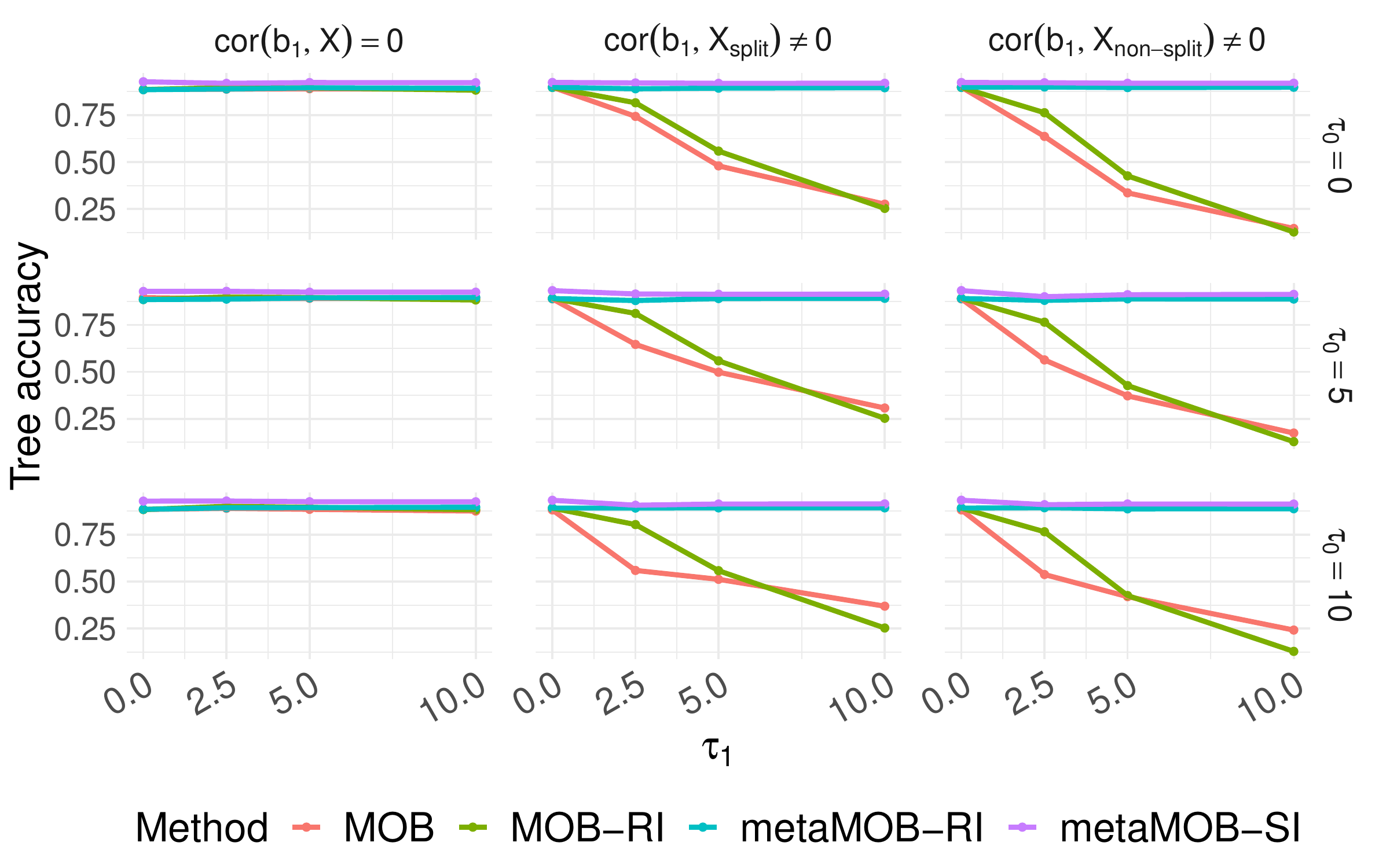}
	\caption{Tree accuracy for Sim A (\ref{M2} as data-generating mechanism) with $b_0$ and the covariates $\mathbf{X}$ are uncorrelated. The correlation of $b_1$ and the covariates $\mathbf{X}$ is varied (columns). Different variances of the random intercept are presented in the three rows of the figure. }
	\label{fig:m1_treeaccuracy_b0uncorr}
\end{figure}

In settings using Sim A as data generating model, the accuracy of MOB-RI and MOB deteriorates with increasing heterogeneity in the treatment effect if the heterogeneity of the treatment effect is correlated with one of the splitting candidates (see Figure \ref{fig:m1_treeaccuracy_b0uncorr}). The accuracy of the trees obtained by MOB or MOB-RI is slightly larger in settings in which the treatment heterogeneity is correlated with $X_1, X_2$ or $X_5$, one of the subgroup-defining covariates, compared to settings in which the treatment heterogeneity is correlated with a non-splitting variable. Without $b_1$ and $\mathbf{X}$ being correlated (see first column of Figure \ref{fig:m1_treeaccuracy_b0uncorr}), no difference between the four methods with regard to tree accuracies is observable. For uncorrelated $b_1$ and $\mathbf{X}$, but $\mathbf{X}$ and  $b_0$ correlated, accuracies much smaller than one are only observable for for MOB. MOB-RI adjusts for between trial heterogeneity in baseline characteristics using a random intercept. Therefore, smaller accuracies for MOB-RI are only observable in settings with $\tau_1^2\neq 0$ and $\mathbf{X}$ and  $b_0$ correlated (results not shown).
The deterioration of the accuracy with increasing variance of the treatment heterogeneity using MOB and MOB-RI is mainly due to estimated trees with too many splits. It seems that MOB and MOB-RI, which do not adjust for heterogeneity in the treatment effect, try to capture this heterogeneity which is linked to covariates by further splitting the data. However, the additional splits are not necessarily performed on covariates correlated to the between-trial heterogeneity. 

\begin{figure}
	\centering
	\includegraphics[width=1\linewidth]{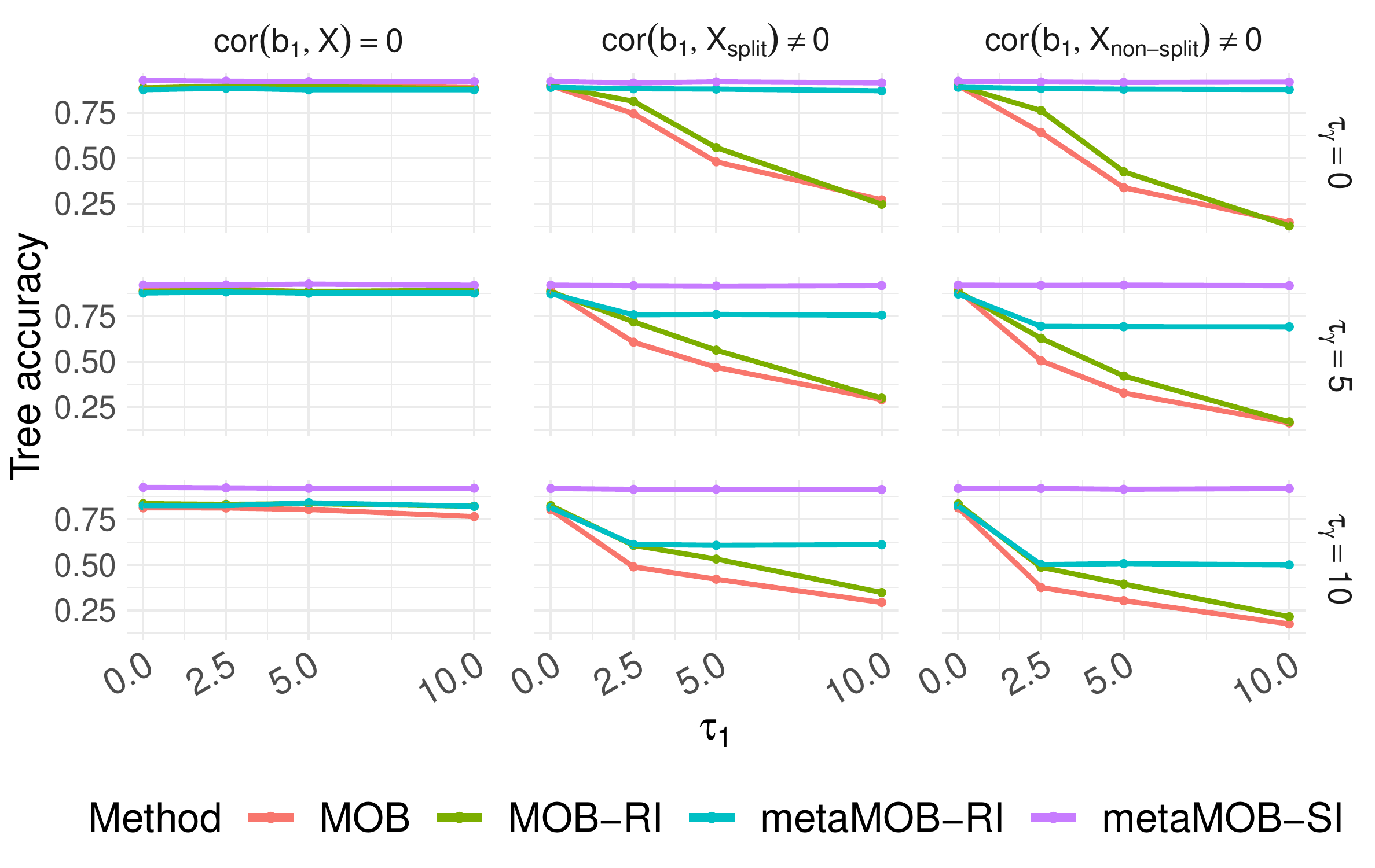}
	\caption{Tree accuracy for Sim B (\ref{M3} as data-generating mechanism). The correlation of $b_1$ and the covariates $\mathbf{X}$ is varied (columns). Different variances for the subgroup and trial specific intercepts are presented in the three rows.}
	\label{fig:m2_treeaccuracy}
\end{figure}

Using \ref{M3} as data generating model as is Sim B, we observe that metaMOB-SI which is using the correct model identifies trees closest to the true underlying tree (see Figure \ref{fig:m2_treeaccuracy}). However, in settings without between-trial heterogeneity in the treatment effect or without correlation of the random treatment effect and one of the potential splitting covariates, the mean accuracies of MOB, MOB-RI and metaMOB-RI are not much smaller than the tree accuracy of metaMOB-SI.
Larger variations in trial and subgroup specific intercepts as the presence of between-trial heterogeneity in the treatment effect result in smaller tree accuracies of MOB, MOB-RI and metaMOB-RI due to their model misspecification. Although metaMOB-RI accounts for heterogeneity in both, the baseline and the treatment effect, the assumption for the baseline is not flexible enough for this setting. The underlying model of metaMOB-RI assumes that the between-trail heterogeneity of the baseline effects is the same across all subgroups, which is not the case in Sim B.

\subsubsection{Correlation of the estimated and true individual treatment effect}
\begin{figure}
	\centering
	\includegraphics[width=1\linewidth]{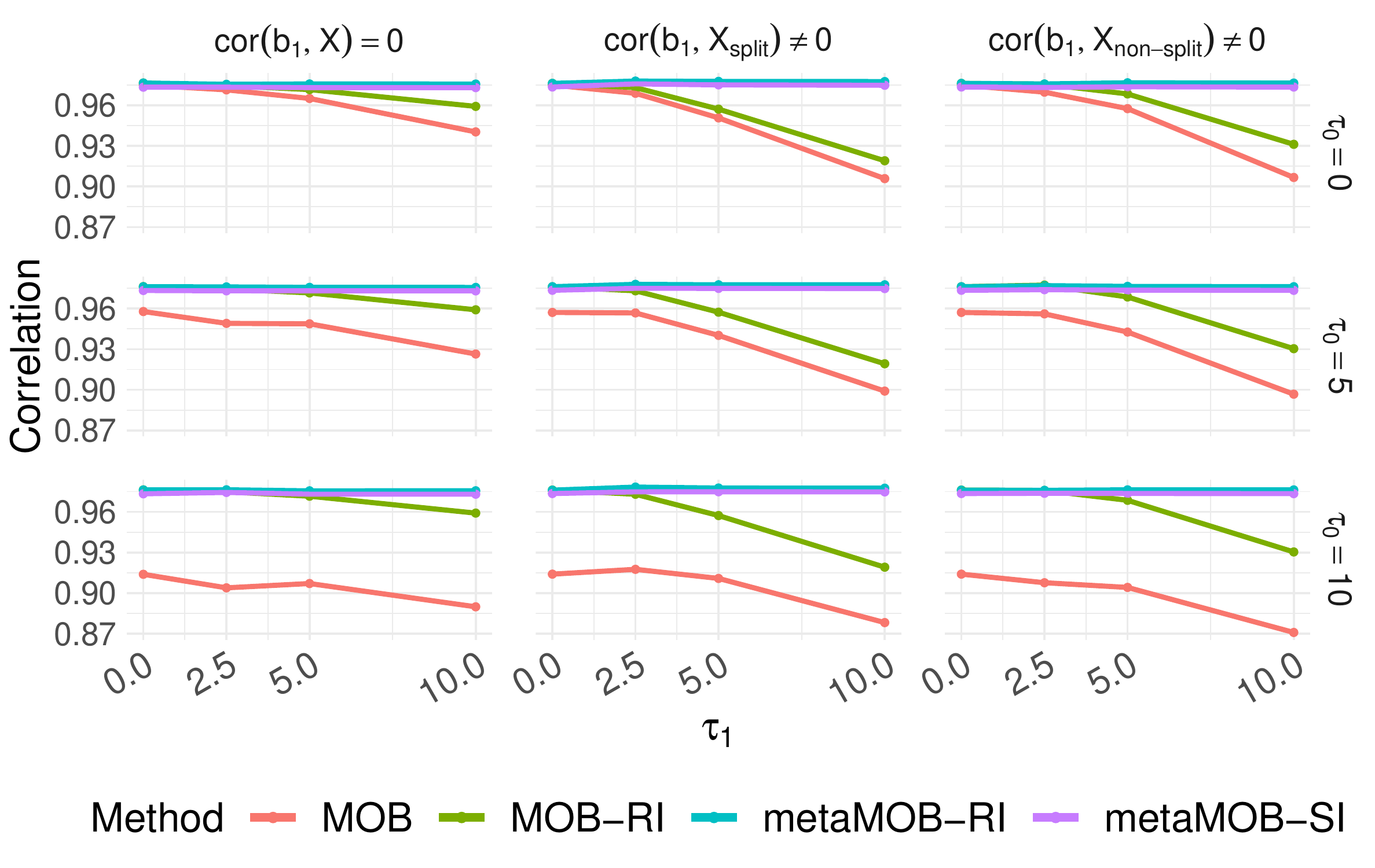}
	\caption{Correlation of true and estimated treatment effect identified subgroups in Sim A. Figure includes only settings in which $b_0$ and the covariates are not correlated.}
	\label{fig:m1courubi0cortrttest}
\end{figure}

For the correlation of the estimated and true individual treatment effect the treatment effects are calculated on a test sample. For this calculation, the patients of the test data are assigned to a subgroup according to the grown tree. Based on this subgroup assignment model \ref{M0} to model \ref{M3} are fitted. The model is chosen according to the MOB algorithm used for growing the tree. The true treatment effect in the estimated subgroups is calculated by a weighted mean of the true treatment effect $\theta$. The weights are based on the number correctly assigned patients to each of the true subgroups. Simulations runs which failed to grow a tree, runs with convergence problems of \ref{M1} to \ref{M3} on the test data and runs in which not all identified subgroups are present in the test data are omitted for the calculation of the mean correlations of the settings presented in Figure \ref{fig:m1courubi0cortrttest}.

Figure \ref{fig:m1courubi0cortrttest} shows that MOB and MOB-RI's correlation of true and estimated treatment effect decreases when random effects are misspecified by MOB's underlying model. This is the case in settings with $\tau_1^2\neq 0$ for both MOB and MOB-RI and in setting with $\tau_0^2\neq0$ for MOB only. For both metaMOB-RI and metaMOB-SI, which have more general models underlying the tree algorithm, correlations close to 1 are observable throughout the considered settings shown in Figure \ref{fig:m1courubi0cortrttest}. The results presented in Figure \ref{fig:m1courubi0cortrttest} consider a larger absolute treatment effect difference, e.g. standardized treatment effects $|\theta_1/\sigma_\epsilon|=|\theta_3/\sigma_\epsilon|=1$ with $\sigma_\epsilon=5$. As the probability of identifying the correct treatment-by subgroup interaction decreases with decreasing treatment-by subgroup interaction effects \citep{alemayehu,huber}, we can also expect smaller correlations of true and estimated treatment effects in the identified subgroups similar to the results in \cite{Fokkema2017}. 

\section{Discussion}\label{discussion}
%Summary
We proposed metaMOB, a tree based method, allowing to identify treatment-by subgroup interactions while accounting for between-trial heterogeneity in IPD- meta-analysis settings. The method metaMOB falls into the broad class of GLMM-trees introduced by \cite{Fokkema2017}. GLMM-trees allow to estimate random effect parameters and to apply recursive partitioning to the data as proposed by \cite{mobZ}. The approach used in the algorithm for GLMM-trees was also adapted in PALM-trees \cite{Seibold2018} which allows to estimate further fixed effects which are constant over all groups instead of random effects.
GLMM-trees were only investigated for a simpler mixed model, namely a random intercept model, underlying the partitioning. We referred to the investigated GLMM-tree as MOB-RI, model-based recursive partitioning with a random intercept.
In a simulation study, we evaluated the performance of MOB, MOB-RI and metaMOB in IPD meta-analysis settings with continuous response. 
We considered two variations of metaMOB, namely metaMOB-RI and metaMOB-SI. Both methods model between-trial variations of the treatment effect using random effects as commonly done in random-effects meta-analysis. However, metaMOB-RI and metaMOB-SI differ in their assumptions regarding the baseline effects. One of the models assumes the between-trial heterogeneity in the baseline effect to be fixed. Therefore, the method uses a stratified intercept approach (metaMOB-SI), whereas the other model assumes the heterogeneity in the baseline effect to be random as well (metaMOB-RI). 
The simulation study showed that all four considered methods perform similarly well in terms of false discovery rate and identifying correct subgroups, when no between-trial heterogeneity is present or when between-trial heterogeneity is independent from potential splitting candidates and can therefore not be explained by any of the covariates. However, in IPD meta-analysis settings we believe it is reasonable to assume that heterogeneity between trials is linked to one or more covariates. For those settings, the simulation study showed that misspecifying the structure of the between-trial heterogeneity results in less accurate trees and high false discovery rates. The misspecified between-trial heterogeneity also affects the estimated treatment effects in the identified subgroups. The correlation of the estimated and true treatment effect in the identified subgroups decreases when we model the data with simpler models compared to the more complex true model. As assumptions regarding treatment effect heterogeneity between trials and a correlation between heterogeneity and covariates seem reasonable in IPD meta-analysis,
we can conclude that MOB and the MOB-RI are not the best option for subgroup identification in IPD meta-analysis settings.
The method metaMOB-SI showed the best performance with regard to false discovery rate, accuracy of identified subgroups and estimated treatment effect throughout all considered simulation settings.
Compared to metaMOB-RI, metaMOB with stratified intercepts has also less convergence problems as only one random effect has to be estimated. 
Commonly, one of the main interest in random-effects meta analysis is obtaining an unbiased summarized treatment effect based on several studies by accounting for between-trial heterogeneity, which is done by both metaMOB-RI and metaMOB-SI. However, metaMOB-SI's underlying model does not impose a constraint on the baseline effects as it is done by metaMOB-RI's underlying model. Therefore, metaMOB-SI is more flexible. It has to be kept in mind, however, that the number of parameters which have to be estimated for metaMOB-SI increases with increasing number of trials and also with increasing number of identified subgroups. %This might cause estimation problems.% if the sample size is not large enough. 
This might lead to inconsistent estimators as described by the Neyman-Scott problem \citep{Neyman1948}.
Therefore, the number of subgroups which can be identified by metaMOB-SI is strongly restricted for meta-analysis with smaller sample sizes and numerous included studies.\\ 
One limitation of our simulation study is that we assumed larger number of trials as motivated by the examples (presented in the Introduction). \cite{Turner2012} found that 50\% of 14886 aggregated meta-analyses included in their analysis were based on 3 trials or less. Only 25\% of the analysed meta-analysis included 6 or more studies. For IPD meta-analysis it might be even harder to retrieve data from a larger number of trials. 
First, larger number of trials have to be eligible for the IPD meta-analysis and second all authors have to agree sharing their data. Contacting and communicating with all authors is time-consuming and might often not be successful due to problems in communication and in data-sharing agreements. Furthermore, the anonymisation of raw data might often be challenging especially in the case of studies with numerous covariates. Although there is an increasing interest in improving reproducible research and therefore making raw data available for further research in practise major hurdles have to be overcome.\\
Difficulties in estimating the between-trial variance may arise, when data from few trials are analysed. Especially higher fractions of estimates of the between-trial heterogeneity in the treatment effect ($\tau_1$) equal to zero in meta-analyses with few trials \citep{Friede2017} may negatively affect subgroup identification using MOB-RI or metaMOB. Further research on the performance of metaMOB for IPD-meta-analysis settings  with few trials is needed.\\
Although we considered continuous outcomes only in our simulation study, the proposed approach is applicable more widely including binary endpoints, since the implementation by \cite{Fokkema2017} is based on generalized mixed effect models. Therefore, metaMOB might also be applicable to discretely measured time-to-event outcomes as the log-likelihood of a discrete time-to-event model and a regression model for a binary outcome are equivalent \citep{Schmid2016}. However, the assessment of the properties of metaMOB with discrete time-to-event endpoints is subject to future research.
\\
SIDES for IPD meta-analysis \citep{Mistry2018} is similar to MOB-RI regarding the assumptions of the between-trial heterogeneity. Both account for between-trtial heterogeneity in the baseline only. Both, SIDES for IPD meta-analysis
and the GLMM-tree framework, however, do not account for between-trial heterogeneity in the treatment effect within identified subgroups. It is assumed that the between-trial heterogeneity accounted for by random effects is constant across identified subgroups. As between-trial variations of the treatment-by subgroup interactions are reasonable assumptions research for identifying subgroups in such settings using tree-based methods is needed.

\section*{Acknowledgement}
This work was cofunded by the BfArM and the University Medical Center G\"ottingen and is a result of the collaborative project ``Identifikation und Konfirmation von Biomarker-definierten Populationen in der personalisierten Pharmakotherapie'' (``Identification and confirmation of biomarker-defined populations in the personalized pharmacotherapy'').
\section*{Author Note}
The views expressed in this article are the authors personal views and do not necessarily represent those of the Federal Institute for Drugs and Medical Devices (BfArM).
\bibliographystyle{spr-chicago}      
\bibliography{lit}
\end{document}